\documentclass[fleqn,twoside]{article}
\usepackage[headings]{espcrc2}
\usepackage{psfrag,epsfig,axodraw,amsmath,feynarts}
\usepackage[dvips]{color}
\usepackage{colordvi}

\title{Regularization of supersymmetric theories --- recent progress}

\author{Dominik St\"ockinger\address{
Institute for Particle Physics 
Phenomenology, University of Durham,\\
Durham DH1~3LE, UK}\thanks{email: Dominik.Stockinger@durham.ac.uk}}

\begin{document}

\begin{abstract}

General issues concerning the regularization of supersymmetric
theories using dimensional regularization and dimensional reduction
are reviewed. Recent progress on problems of dimensional reduction
related to factorization, supersymmetry, Higgs boson mass
calculations, and mathematical consistency is discussed.

\end{abstract}

\maketitle

\section{Introduction}

Regularization is a necessary step in any computation of radiative
corrections or quantum effects in quantum field theory. Its purpose is
to provide an intermediate definition of otherwise divergent (loop or
phase space) integrals. In principle any regularization scheme can be
chosen as long as it is consistent with fundamental properties like
unitarity and causality. However, in practice a wise choice
is important in order to simplify  the evaluation of the integrals as
well as of the counterterms necessary for the renormalization
procedure. 

In practice, the most common schemes for perturbative calculations are
dimensional regularization (DREG) \cite{HV} and dimensional reduction
(DRED) \cite{Siegel79}. Both allow for very efficient techniques for
evaluating momentum integrals. In spite of being the best known
options, however, both schemes lead to well-known problems when
applied to supersymmetric theories.

In DREG not only momentum integrals are continued to $D$
  dimensions, but also gauge fields are treated as $D$-component
  quantities. The mismatch between the number of degrees of freedom of
  gauge fields ($D$) and gauginos (4) breaks supersymmetry in
  DREG. Therefore, special supersymmetry-restoring counterterms have
  to be added in order to restore supersymmetry in the renormalized
  theory. These counterterms 
  do not originate from the original Lagrangian by multiplicative
  renormalization, and hence their structure and evaluation poses a
  significant complication of the calculation (for examples see e.g.\
  \cite{BHZ96,STIChecks}).

DRED is better suited to supersymmetry because gauge fields remain
  4-component quantities. In many examples
  DRED has explicitly been found to preserve supersymmetry
  \cite{BHZ96,STIChecks,CJN80}. Nevertheless, for a long time DRED
  has been known to have several, more
  subtle problems, which are the main focus of the present
  talk.%
\footnote{%
Some of these problems have also been reviewed in
\cite{JJ} and highlighted in the ``Supersymmetry Analysis Project'' 
\cite{SPA}.
}
 In
  brief, these problems are the following:
\begin{enumerate}
\item The mathematical formulation of DRED is
  plagued by inconsistencies uncovered in \cite{Siegel80}. As a
  result, certain initial expressions can lead to several different
  results depending on the order of the calculational steps. 
\item It is unknown to what extent DRED actually preserves
  supersymmetry. A general proof that supersymmetry is preserved in all
  cases doesn't exist due to point 1. In fact  
  DRED will break supersymmetry in the gauge boson/gaugino sector
  at least at the 4-loop level \cite{ACV}. The important
  question is whether DRED preserves supersymmetry at least in cases
  that are of phenomenological importance.
\item An apparent inconsistency of DRED with QCD-factorization has
  been observed in \cite{BKNS}. Although factorization seems to be
  a regularization-independent property and general formalisms
  have been worked out both for DREG and DRED (at least in the case
  with massless partons) \cite{CST}, the factorization-problem of
  DRED has repeatedly appeared in the literature
  \cite{BHZ96,vNS}. In these references, DRED has been
  abandoned as a regularization for hadron processes, even in the
  presence of supersymmetry. The question is whether and how it is
  possible to reconcile DRED with factorization.
\end{enumerate}
The following sections will describe recent progress on these
questions.

\section{Consistent DRED and its properties}

It turns out that it is possible to reformulate DRED in a way that
avoids mathematical inconsistencies, i.e.\ such that any initial
expression leads to a unique final result \cite{DREDPaper} (see also
\cite{DREDProc}). This reformulation is not different from the
standard definition of DRED as long as no purely 4-dimensional
identities such as Fierz rearrangements are used. Hence, in most
practical applications of DRED (in particular in all applications
referred to here) actually the consistent version of DRED has been
applied. 

The important practical consequence of having a mathematically
consistent definition is that general properties can be proven. In
particular, the quantum action principle can be established
for DRED \cite{DREDPaper}. This is a general relation between symmetry
properties of Green functions in a regularized quantum field theory
and the corresponding symmetry properties of the Lagrangian:
\begin{align}
i\,\delta\langle T\phi_1\ldots\phi_n\rangle
&= \langle T\phi_1\ldots\phi_n\Delta\rangle\ ,
\label{QAP}
\\
\Delta&=\int d^Dx \,\delta{\cal L}.
\end{align}
Here $\delta$ corresponds to any symmetry transformation, e.g.\
gauge/BRS transformations, supersymmetry transformations etc.
This quantum action principle, established long ago in the
context of BPHZ renormalization \cite{QAPBPHZ} and of DREG \cite{BM},
is of great use in order to study symmetry-properties of
regularization schemes.

In order to give an example of the direct relevance of the quantum
action principle, consider non-supersymmetric QCD and DREG and take
$\delta$ to represent BRS transformations. It is one of the most
beautiful features of DREG that it doesn't modify the structure of the
QCD Lagrangian, and therefore the Lagrangian is BRS invariant even on
the regularized level, $\delta_{\rm BRS}{\cal L}_{\rm QCD}=0$. Hence
the quantum action principle for DREG shows that
\begin{align}
\Delta_{\rm BRS}=0\Rightarrow
\delta_{\rm BRS}\langle T\phi_1\ldots\phi_n\rangle=0.
\end{align}
The right equation here is nothing but a generic QCD Slavnov-Taylor
identity. In this way, we find that in DREG all QCD Slavnov-Taylor
identities are valid even on the regularized level (and at all
orders) --- a statement of utmost practical importance.

Since the quantum action principle has now been shown to be valid also
in DRED we can derive the supersymmetry-properties of DRED in an
analogous fashion. The result for the operator $\Delta_{\rm SUSY}$
corresponding 
to the supersymmetry variation of the Lagrangian of a generic
supersymmetric model in DRED has been given in \cite{DREDPaper}. In
contrast to the QCD/BRS case it does not vanish. Therefore, applying
the quantum action principle does not lead to the statement that DRED
preserves all supersymmetry identities at all orders. On the contrary,
one has to expect that DRED will violate supersymmetry at some point,
as already argued in \cite{ACV}. In general, the quantum action
principle yields the following equivalence: a given supersymmetry
identity
\begin{align}
\delta_{\rm SUSY}\langle T\phi_1\ldots\phi_n\rangle=0
\label{GenericSUSYId}
\end{align}
is valid in DRED on the regularized level if and only if 
\begin{align}
\langle T\phi_1\ldots\phi_n\Delta_{\rm SUSY}\rangle=0.
\label{GenericDelta}
\end{align}
Often in practice, (\ref{GenericDelta}) is a lot easier to verify than
(\ref{GenericSUSYId}). 

\section{DRED and supersymmetry}
\label{sec:susy}

As mentioned in the Introduction, it is important to know to what
extent DRED preserves supersymmetry. The reason is that only
if DRED preserves supersymmetry (in a given sector/loop order) the
familiar concept of multiplicative renormalization is correct, where
``multiplicative renormalization'' is meant as a synonym for any of
the following:
\begin{itemize}
\item Counterterms are generated by multiplicative renormalization of
  the parameters and fields of the Lagrangian.
\item The bare Lagrangian has the same structure as the renormalized
  Lagrangian (which generates the Feynman rules).
\item The bare Lagrangian and the counterterm Lagrangian have the
  same symmetries as the renormalized Lagrangian.
\end{itemize}
If DRED violates supersymmetry, all of this is not true and
additional, supersymmetry-restoring counterterms have to be found and
added. The existence of such counterterms is guaranteed by the absence
of anomalies and the renormalizability of supersymmetric models
\cite{ren}, but their necessity complicates practical calculations.
So far, in virtually all practical applications of DRED, DRED has been
assumed to preserve supersymmetry, however often without explicit
investigation or proof.

Until recently, the only explicit checks of supersymmetry in DRED
concerned one-loop relations for 2-point \cite{CJN80} and 3-point
functions \cite{BHZ96,STIChecks} and relations between
renormalization group $\beta$-functions \cite{betachecks}. These
checks were done by explicitly evaluating all Green functions
appearing in the corresponding supersymmetry identity (i.e.\ the
left-hand side of eq.\ (\ref{GenericSUSYId})). They suffice to prove
that e.g.\ for one-loop calculations of 
processes such as sfermion-pair or chargino-pair production DRED
preserves supersymmetry and multiplicative renormalization is
correct. 

Using the quantum action principle, supersymmetry identities can be
checked much more easily by explicitly evaluating the left-hand side
of eq.\ (\ref{GenericDelta}). In this way, already several 2-loop
identities could be shown to be preserved by DRED \cite{DREDPaper}.

In the following we consider the theoretical evaluation of the
lightest Higgs boson mass in the 
minimal supersymmetric standard model (MSSM) as a
particularly prominent case where DRED has been assumed to preserve
supersymmetry in the literature (see the review \cite{HHW} and
references therein). 

\begin{figure}[tb]
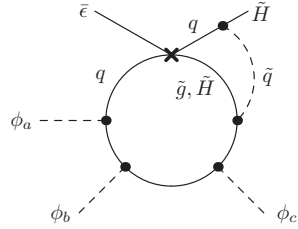

\begin{center}
\scalebox{.8}{
\unitlength=1.cm%
\begin{feynartspicture}(4,4)(1,1)
\FADiagram{}
\FAVert(14.66,18.66){0}
\FAVert(16,10){0}
\FAVert(10,16){1}
\FALabel(3,20)[r]{$\bar\epsilon$\ }
\FAProp(17,20)(10,16)(0.,){/Straight}{0}
\FALabel(17,20)[l]{\,$\tilde{H}$}
\FAProp(3,20)(10,16)(0.,){/Straight}{0}
\FALabel(4,14)[r]{$q$}
\FALabel(18,14)[l]{\,$\tilde{q}$}
\FALabel(14,13)[r]{$\tilde{g},\tilde{H}$}
\FALabel(12,18)[b]{${q}$}
\FAProp(10,16)(10,4)(1.,){/Straight}{0}
\FAProp(10,16)(10,4)(-1.,){/Straight}{0}
\FAProp(14.66,18.66)(16,10)(-.5,){/ScalarDash}{0}
\FAProp(-2,10)(4,10)(0.,){/ScalarDash}{0}
\FAProp(5.76,5.76)(1.51,1.51)(0.,){/ScalarDash}{0}
\FAProp(14.24,5.76)(18.48,1.51)(0.,){/ScalarDash}{0}
\FALabel(19,1.5)[l]{\ $\phi_c$}
\FALabel(1.5,1.5)[r]{ $\phi_b$\ }
\FALabel(-2,10)[r]{ $\phi_a$\ }
\FAVert(5.76,5.76){0}
\FAVert(14.24,5.76){0}
\FAVert(4,10){0}
\end{feynartspicture}
}
\end{center}
\vspace{-1.3cm}
\caption{A particular diagram contributing to the Green function in
  eq.\ (\ref{DeltaGreenFunction}). The cross represents the insertion
  of the operator $\Delta_{\rm SUSY}$. $q$, $\tilde{q}$, and
  $\tilde{g}$ denote quark, squark, and gluino lines. 
  $\bar\epsilon$ is the supersymmetry transformation parameter.
According to the general rules derived in \cite{DREDPaper} this
  diagram vanishes since the fermion loop contains less than four
  $\gamma$-matrices.
}
\label{fig:dangerous}
\end{figure}

Generically, the Higgs boson mass is related to the quartic Higgs boson
self coupling $\lambda$ as
\begin{align}
M_h^2\propto \lambda v^2
\end{align}
where $v$ is the vacuum expectation value. In supersymmetric models
such as the MSSM, $\lambda$ is not a free parameter, but it is related
to gauge couplings,
\begin{align}
\lambda\propto g^2.
\label{SUSYlambda}
\end{align}
Therefore, supersymmetry determines the Higgs boson mass, and precise
theoretical predictions taking into account up to two-loop corrections
are very important. The relation (\ref{SUSYlambda}) is the crucial
supersymmetry-relation that has been assumed to be preserved by DRED
in all evaluations of $M_h$ described in \cite{HHW}.

Recently, this assumption has been explicitly checked
\cite{MhPaper}. The check has three main elements. 
Firstly, the
supersymmetry relation (\ref{SUSYlambda}) is formulated on the
level of Green functions as a Slavnov-Taylor identity,
\begin{align}
\delta_{\rm SUSY}\langle T\phi_a\phi_b\phi_c
\tilde{H}\rangle =0,
\label{STICheck}
\end{align}
where $\phi_{a,b,c}$ and $\tilde{H}$ denote Higgs and Higgsino
fields (for a more explicit and detailed form see \cite{MhPaper}).
Secondly, this identity between DRED-regularized Green functions is
replaced by the equivalent relation
\begin{align}
\langle T\phi_a\phi_b\phi_c\tilde{H}\Delta_{\rm SUSY}\rangle=0
\label{DeltaGreenFunction}
\end{align}
by virtue of the quantum action principle (\ref{GenericSUSYId}),
(\ref{GenericDelta}). Finally, the Green function 
$\langle T\phi_a\phi_b\phi_c\tilde{H}\Delta_{\rm SUSY}\rangle$
is evaluated.

The level to which the Green function is evaluated is the level of
two-loop Yukawa-enhanced contributions, i.e. contributions of ${\cal
  O}(\alpha_{t,b}\alpha_s)$, ${\cal O}(\alpha_{t,b}^2)$,  
${\cal O}(\alpha_t\alpha_b)$. This is the order of current state-of-the-art
computations of the MSSM Higgs boson mass \cite{HHW}.
Fig.\ \ref{fig:dangerous} shows one of the diagrams that have to be
evaluated. It turns out that the Feynman rules corresponding to the
insertion $\Delta_{\rm SUSY}$ lead to rather simple rules as to when
such a diagram vanishes. In cases like the one shown in Fig.\
\ref{fig:dangerous} where one closed fermion loop and one outgoing
quark line are attached to $\Delta_{\rm SUSY}$, the diagram vanishes
if the number of $\gamma$-matrices in the fermion loop is smaller than
four. Inspection of the diagram in Fig.\ \ref{fig:dangerous} shows
that indeed the fermion loop contains only terms with less than four
$\gamma$-matrices, and therefore the diagram vanishes. In a similar
way it is straightforward to show that all diagrams contributing to
the Green function in eq.\ (\ref{DeltaGreenFunction}) vanish.

This result proves that (\ref{DeltaGreenFunction}) and hence also
(\ref{STICheck}) are valid in DRED at ${\cal
  O}(\alpha_{t,b}\alpha_s)$, ${\cal O}(\alpha_{t,b}^2)$,  
${\cal O}(\alpha_t\alpha_b)$. 

\section{DRED and factorization}

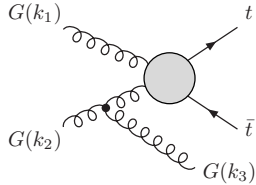
\begin{figure}
\centerline{
\scalebox{.8}{
\begin{picture}(130,65)(0,20)
\SetWidth{.6}
\put(-12,75){$G(k_1)$}
\put(-12,15){$G(k_2)$}
\put(80,0){$G(k_3)$}
\put(100,75){$t$}
\put(100,20){$\bar t$}
\Gluon(14,70)(54,52){3}{5.5} 
\Gluon(34,33)(54,42){3}{2}
\Gluon(14,24)(34,33){3}{2}
\Gluon(76,5)(34,33){3}{6}
\Vertex(34,33){2}
\ArrowLine(72,55)(96,71)
\ArrowLine(96,23)(72,39)
\GCirc(64,47){12}{.85}
\end{picture}
}
}
\caption{
\label{fig:diagrams}
Generic structure of NLO diagrams $GG\to t\bar t G$ giving rise to a collinear divergence for
$k_3\to(1-x)k_2$.}
\end{figure}

The factorization-problem of DRED mentioned in the Introduction has
been first observed in the evaluation of the next-to-leading order
(NLO) corrections to the process $GG\to t\bar t$ \cite{BKNS}. The real
NLO corrections from diagrams with an additional unresolved
final-state gluon exhibit collinear divergences from the diagrams
shown in Fig.\ \ref{fig:diagrams} if gluons 2 and 3 become parallel.
However, contrary to expectations and experience from calculations
in DREG, the collinear limit does not factorize. Instead, the
divergent part of the NLO cross section $\sigma_{GG\to t\bar t G}$
is proportional to
\begin{align}
P_{G\to GG}\ 
\sigma_{GG\to t\bar t}
+\frac{4-D}{D-2}\ 
\frac{(1-x)}{x}\ \tilde{\sigma}.
\label{CollDRED}
\end{align}
The appearance of the LO cross section, multiplied by the splitting
function $P_{G\to GG}$ is expected, but the appearance of
the additional term $\tilde{\sigma}$, which is not proportional to the 
LO cross section, is puzzling. In DREG, the non-factorizing term is
absent. 

The question therefore is whether DRED has to be abandoned as a
regularization of hadron processes or whether DRED can be reconciled
with factorization. These questions have been addressed in
\cite{FacPaper}.

The key to understanding what happens in eq.\ (\ref{CollDRED}) is to
realize that in DRED there is a mismatch between the number of gluon
components (4) and the number of dimensions ($D$). As a result, on the
regularized level the 4-component gluon $G$ splits into a
$D$-component part, denoted by $g$, and the remaining $4-D$ components
$\phi$. Only $g$ acts as the $D$-dimensional gauge field; the fields
$\phi$ are simply scalar fields, called $\epsilon$-scalars
\cite{CJN80}.

It is natural to treat $g$ and $\phi$ as two independent partons. Then
diagrams of the form shown in Fig.\ \ref{fig:diagrams} with an
intermediate 4-component gluon $G$ can be decomposed into two
contributions: one contribution with a virtual $g$ and another one
with a virtual $\phi$. Using this decomposition it is possible to rewrite the
collinear limit (\ref{CollDRED}) in the following form \cite{FacPaper}:
\begin{align}
P_{G\to gG}
\sigma_{Gg\to t\bar t}
+
P_{G\to \phi G}
\sigma_{G\phi\to t\bar t}.
\label{newDREDexpectation}
\end{align}
This form of the collinear limit is fully in agreement with
factorization in a theory that has two independent partons $g$,
$\phi$. 

It is noteworthy that the puzzling non-factorizing term
$\tilde{\sigma}$ can be written as
\begin{align}
\tilde\sigma \propto 
(\sigma_{Gg\to t\bar t}
-\sigma_{G\phi\to t\bar t})
.
\end{align}
Hence it is obvious that the puzzling term only appears since the two
partonic cross sections involving $g$ and $\phi$ are different. In
some kinematically simpler processes such as $GG\to q\bar q$ with
a massless quark $q$ these cross sections are identical, 
i.e.\ $\sigma_{Gg\to q\bar q}=\sigma_{G\phi\to q\bar q}$. This
explains why the factorization-problem of DRED has only appeared in
kinematically complicated processes such as pair production of massive
quarks, squarks, or gluinos.

In general, however, the result (\ref{newDREDexpectation}) demonstrates that
DRED can be reconciled with factorization. As explained in
\cite{FacPaper}, this implies that there is no obstacle to evaluate
hadron processes using DRED.

\section{Conclusions}

After recent progress in \cite{DREDPaper,MhPaper,FacPaper} the
status of DRED is quite satisfactory. DRED is consistent with
factorization, and DRED has been proven to preserve supersymmetry in
a large range of cases, including 2-loop relations relevant for the
MSSM Higgs boson mass. Further studies of both aspects will still be
important. In this respect it is promising that the validity of the
quantum action principle makes feasible many future checks of
supersymmetry-properties of DRED.

\paragraph{Acknowledgments}

I want to thank the organizers and the participants of RADCOR05 for
creating an interesting and stimulating conference.


\begin{thebibliography}{99} 
\bibitem{HV} G.~'t Hooft and M.~Veltman,
               {\em Nucl. Phys.} {\bf B 44} (1972) 189.
\bibitem{Siegel79} W.~Siegel,
                       {\em Phys. Lett.} {\bf B 84} (1979) 193.



\bibitem{BHZ96}
W.~Beenakker, R.~H\"opker and P.~M.~Zerwas,
{\em Phys.\ Lett.}{\bf\ B} {\bf 378} (1996) 159.


\bibitem{STIChecks}
W.~Hollik, E.~Kraus and D.~St\"ockinger,
{\em Eur.\ Phys.\ J.}{\bf\ C} {\bf 11} (1999) 365;
W.~Hollik and D.~St\"ockinger,
{\em Eur.\ Phys.\ J.}{\bf\ C} {\bf 20} (2001) 105;
I.~Fischer, W.~Hollik, M.~Roth and D.~St\"ockinger,
{\em Phys.\ Rev.}{\bf\ D} {\bf 69} (2004) 015004.

\bibitem{CJN80}
D.~M.~Capper, D.~R.~T.~Jones and P.~van Nieuwenhuizen,
{\em Nucl.\ Phys.}{\bf\ B} {\bf 167} (1980) 479.

\bibitem{JJ}
I.~Jack and D.~R.~T.~Jones,
[arXiv:hep-ph/9707278].

\bibitem{SPA}
J.~A.~Aguilar-Saavedra {\it et al.},
arXiv:hep-ph/0511344.


\bibitem{Siegel80}
W.~Siegel,
{\em Phys.\ Lett.}{\bf\ B }{\bf 94} (1980) 37.

\bibitem{ACV}
L.~V.~Avdeev, G.~A.~Chochia and A.~A.~Vladimirov,
{\em Phys.\ Lett.}{\bf\ B} {\bf 105} (1981) 272;
L.~V.~Avdeev and A.~A.~Vladimirov,
{\em Nucl.\ Phys.}{\bf\ B }{\bf 219} (1983) 262.

\bibitem{BKNS}
W.~Beenakker, H.~Kuijf, W.~L.~van Neerven and J.~Smith,
{\em Phys.\ Rev.}{\bf\ D} {\bf 40} (1989) 54.

\bibitem{CST}
S.~Catani, M.~H.~Seymour and Z.~Trocsanyi,
{\em Phys.\ Rev.}{\bf\ D} {\bf 55} (1997) 6819.


\bibitem{vNS}
J.~Smith and W.~L.~van Neerven,
{\em Eur.\ Phys.\ J.}\  {\bf C 40} (2005) 199.


\bibitem{DREDPaper}
D.~St\"ockinger,
{\em JHEP} {\bf 0503} (2005) 076.

\bibitem{DREDProc}
D.~St\"ockinger,
arXiv:hep-ph/0506258.

\bibitem{QAPBPHZ}
J.~H.~Lowenstein,
{\em Phys.\ Rev.}{\bf\ D} {\bf 4} (1971) 2281,
{\em Commun.\ Math.\ Phys.}\  {\bf 24} (1971) 1;
Y.~M.~Lam,
{\em Phys.\ Rev.}{\bf\ D} {\bf 6} (1972) 2145;
{\em Phys.\ Rev.}{\bf\ D} {\bf 7} (1973) 2943.

\bibitem{BM}   P.~Breitenlohner and D.~Maison,
               {\em Commun. Math. Phys.} {\bf 52} (1977) 11.


\bibitem{ren} P.~L.~White,
{\em Class.\ Quant.\ Grav.}\  {\bf 9} (1992) 1663;
N.~Maggiore, O.~Piguet and S.~Wolf,
{\em Nucl.\ Phys.}{\bf\ B }{\bf 458} (1996) 403
[Erratum-ibid.\ B {\bf 469} (1996) 513],
{\em Nucl.\ Phys.}{\bf\ B }{\bf 476} (1996) 329;
W.~Hollik, E.~Kraus and D.~St\"ockinger,
{\em Eur.\ Phys.\ J.}{\bf\ C }{\bf 23} (2002) 735;
W.~Hollik, E.~Kraus, M.~Roth, C.~Rupp, K.~Sibold and D.~St\"ockinger,
{\em Nucl.\ Phys.}{\bf\ B} {\bf 639} (2002) 3.

\bibitem{betachecks}
S.~P.~Martin and M.~T.~Vaughn,
{\em Phys.\ Lett.}{\bf\ B} {\bf 318} (1993) 331;
I.~Jack, D.~R.~T.~Jones and C.~G.~North,
{\em Nucl.\ Phys.}{\bf\ B} {\bf 473} (1996) 308,
{\em Phys.\ Lett.}{\bf\ B} {\bf 386} (1996) 138,
{\em Nucl.\ Phys.}{\bf\ B} {\bf 486} (1997) 479.

\bibitem{HHW}
S.~Heinemeyer, W.~Hollik and G.~Weiglein,
arXiv:hep-ph/0412214.

\bibitem{MhPaper}
W.~Hollik and D.~St\"ockinger,
hep-ph/0509298.

\bibitem{FacPaper}
A.~Signer and D.~St\"ockinger,
{\em Phys.\ Lett.}{\bf\ B} {\bf 626} (2005) 127.

\end{thebibliography}
\end{document}